\documentclass{article}

\usepackage{arxiv}

\usepackage[utf8]{inputenc} 
\usepackage[T1]{fontenc}    
\usepackage{hyperref}       
\usepackage{url}            
\usepackage{booktabs}       
\usepackage{amsfonts}       
\usepackage{nicefrac}       
\usepackage{microtype}      
\usepackage{lipsum}		
\usepackage{graphicx}
\usepackage[numbers]{natbib}
\usepackage{doi}
\usepackage{caption}
\usepackage{threeparttable}
\usepackage{placeins}
\usepackage{float} 
\usepackage[affil-it]{authblk} 
\usepackage{hyperref} 

\usepackage{mathtools} 
\usepackage[export]{adjustbox}

\usepackage{caption}
\usepackage{booktabs}

\def\showtables{1}

\title{Improve Cross-Modality Segmentation by Treating T1-weighted MRI Images as Inverted CT Scans}

\author[1,2]{\textbf{Hartmut Häntze}}
\author[1]{Lina Xu}
\author[1]{Maximilian Rattunde}
\author[1]{Leonhard Donle}
\author[1,3]{Felix J. Dorfner}
\author[2]{Alessa Hering}
\author[4]{Lisa C. Adams}
\author[4,5]{\textbf{Keno K. Bressem}\textsuperscript{$\ddagger$}}

\affil[1]{Department of Radiology, Charité - Universitätsmedizin Berlin corporate member of Freie Universität Berlin and Humboldt Universität zu Berlin, Hindenburgdamm 30, 12203 Berlin, Germany}
\affil[2]{Diagnostic Image Analysis Group, Radboud University Medical Center, Geert Grooteplein Zuid 10, 6525 GA Nijmegen, the Netherlands}
\affil[3]{Athinoula A. Martinos Center for Biomedical Imaging, Massachusetts General Hospital and Harvard Medical School, 149 Thirteenth St, Charlestown, MA 02129, USA}
\affil[4]{Department of Diagnostic and Interventional Radiology, School of Medicine and Health, Klinikum rechts der Isar, TUM University Hospital, Technical University of Munich Ismaninger Str. 22, 81675 Munich, Germany}
\affil[5]{Department of Cardiovascular Radiology and Nuclear Medicine, School of Medicine and Health, German Heart Center, TUM University Hospital, Technical University of Munich, Lazarettstr. 36, 80636 Munich}

\date{}

\renewcommand{\headeright}{}
\renewcommand{\shorttitle}{Improve Cross-Modality Segmentation}

\hypersetup{
pdftitle={Improve Cross-Modality Segmentation by Treating T1-weighted MRI Images as Inverted CT Scans},
pdfsubject={MRI, CT, Segmentation, Cross-Modality, Radiology, Renal Cell Carcinoma},
pdfauthor={Hartmut Häntze, Lina Xu, Leonhard Donle, Felix J. Dorfner, Alessa Hering, Lisa C. Adams, Keno K. Bressem},
pdfkeywords={MRI, CT, Segmentation, Cross-Modality, Radiology, Renal Cell Carcinoma},
}

\begin{document}
\maketitle
\textsuperscript{$\dagger$}Corresponding author: keno.bressem@tum.de \\

\abstract{
Computed tomography (CT) segmentation models often contain classes that are not currently supported by magnetic resonance imaging (MRI) segmentation models. In this study, we show that a simple image inversion technique can significantly improve the segmentation quality of CT segmentation models on MRI data.
We demonstrate the feasibility for both a general multi-class and a specific renal carcinoma model for segmenting T1-weighted MRI images. Using this technique, we were able to localize and segment clear cell renal cell carcinoma in T1-weighted MRI scans, using a model that was trained on only CT
data. Image inversion is straightforward to implement and does not require dedicated graphics processing units, thus providing a quick alternative to complex deep modality-transfer models. Our results demonstrate that existing CT models, including pathology models, might be transferable to the MRI
domain with reasonable effort.
}
\endabstract

\section{Introduction}
Medical image segmentation plays an important role in many automated image analysis tools. It has been well established for computed tomography (CT) scans, with multiple open-source models \cite{ct_lungs, totalsegmentator} and challenges available. Segmentation of magnetic resonance imaging (MRI) scans, particularly multi-organ segmentation, has long been lacking behind but recently, MRI models have been published, that close this gap \cite{hantze2024mrsegmentator,d2024totalsegmentator,zhuang2024mrisegmentator}. Nonetheless, differences between MRI and CT segmentation remain. For CT TotalSegmentator \cite{totalsegmentator} can parallelly infer 117 Structures while the MRI segmentation model, with the largest number of classes \cite{zhuang2024mrisegmentator}, is limited to 62 structures. Similarly, for other examples such as kidney tumor delineation, public challenges exist for CT (i.e. KiTS23) but not for MRI.

Training new (MRI) segmentation models requires a large number of annotated images, and the more classes involved, the greater the annotation effort needed. Pixel-wise annotation can therefore be very time-consuming, especially if the segmentations have to be created from scratch. However, if annotators instead focus on refining pre-segmented structures, i.e. segmentations that are not perfect but cover a large part of the target structure, the annotation process can be greatly accelerated.

One strategy to create pre-segmentations consists of retraining a new segmentation model for the target domain. For example, by co-registering corresponding CT and MRI scans \cite{srikrishna2021deep} or by using augmented CT scans \cite{spie}. 
Nonetheless, implementing and training an augmented model is resource-intensive, time-consuming, and technically challenging. Additionally, it requires having data of both the source and target modality.

In this article, we demonstrate that image augmentation, specifically inversion, can be sufficient to bridge the gap between MRI and CT segmentation performance. One key difference between MRI and CT images is that dense tissue, such as bones, appears bright (hyperdense) in CT scans but dark (hypointense) in MRI images. We attempt to minimize this difference by using negatives of MRI images and analyze whether it influences the semantic segmentation performance of models trained solely on CT data. We investigate the effects on both a general multi-class segmentation model, i.e. TotalSegmentator, and a specialized pathological model to segment clear cell renal cell carcinoma (ccRCC).

\section{Methods}
This study was approved by the local ethics committee (EA4/062/20). Due to the retrospective nature of the study, patient consent was waived. 

We investigated the effect of inversion on two models: For the multiclass model we used the fast version of TotalSegmentator including its pretrained weights (v2.2.1). To create a specialized ccRCC model we trained an nnU-Net \cite{isensee2021nnu} on an in-house CT dataset to predict left and right kidneys and primary tumor. The training data consisted of 1012 scans from patients with ccRCC. That includes early venous (n=425), delayed venous (n=121), arterial (n=300) and non-contrast phase (n=166). Two radiologists (LCA, KKB), with 5 and 4 years of experience respectively, annotated primary tumor and the corresponding kidney. The remaining kidney was segmented with MRSegmentator \cite{hantze2024mrsegmentator}. 

To test both models we randomly selected 100 ccRCC patients (m: 50, f: 50) of an in-house dataset with one T1- and one T2-weighted MRI sequence each. The patients do not overlap with the CT training data. All scans depict the abdominal region. A medical student (MR) and the same radiologists annotated the primary tumors, which had a median volume of 29 cm$^3$ (min: 0.7  cm$^3$, max: 1971  cm$^3$). In 45 cases the primary tumor was on the left side and in 55 cases on the right side. Lastly, we created an additional groundtruth of 24 abdominal structures with MRSegmentator. 

\begin{figure}
\centering
\includegraphics[width=0.8\linewidth]{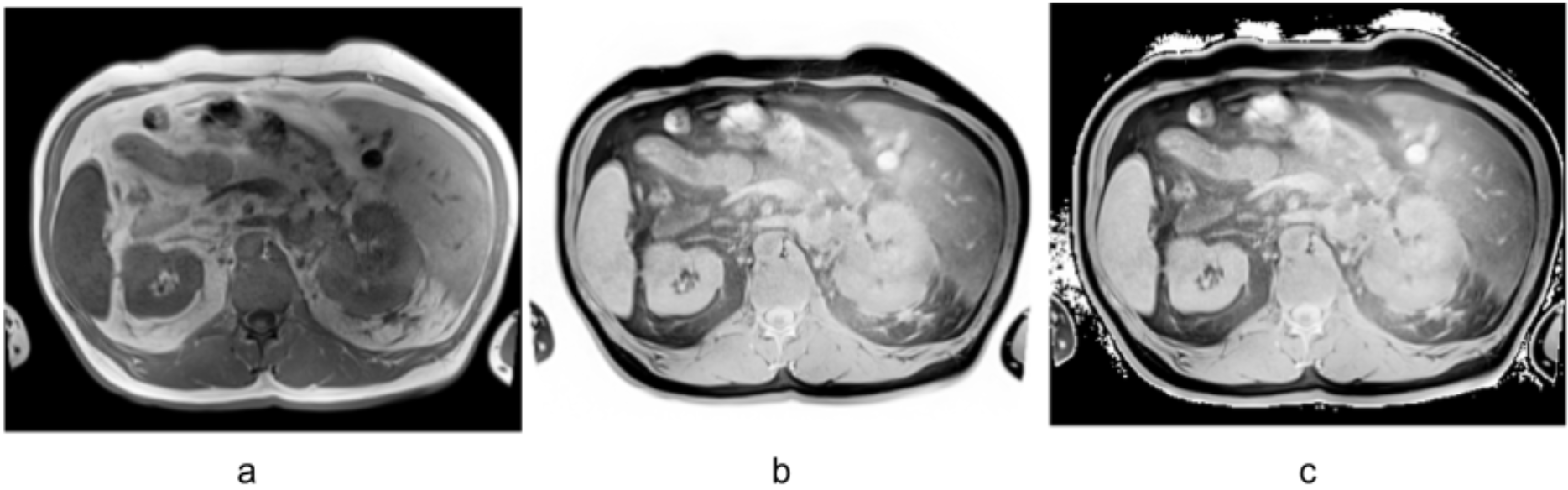}
\caption{(a) Unprocessed T1 image, (b) inverted image, (c) inverted + black background}
\label{fig:preprocessing}
\end{figure}

As preprocessing of the MRI scans we clipped all intensities to a value range between 0 and 3000 and created negatives within their original intensity range. Then, we set all intensities within the first percentile to zero (Equation \ref{eqn:invert}). This step ensures that the surrounding area around the patient remains black (Figure \ref{fig:preprocessing}). Although this process produced some artifacts in the air-filled lungs, it proved to be very stable within the abdominal region. We then ran the TotalSegmentator and our ccRCC model on both the original MRI images and their inverted versions. We compared the model's output to these ground truth labels and calculated the Dice Similarity Coefficient (DSC). Finally, we investigated the role of tumor volume for the segmentation of inverted T1-weighted MRI, specifically.

\begin{equation}
 \label{eqn:invert}   
INV(x) = \begin{dcases*}
0 & x $\in percentile_{1}(X)$ \\
max(X) -x + min(X) & x else
\end{dcases*}
\end{equation}

\section{Results}

\ifdefined\showtables
\begin{table}
\centering
\caption{Average Dice Similarity Coefficient and 95\% Confidence Interval}%
\label{tab:results}

\begin{tabular}{cccc}

  \multicolumn{4}{c}{\bfseries 24 Classes (TotalSegmentator)} \\
  \hline
  \bfseries Sequence & \bfseries Unprocessed  & \bfseries Inverted & \bfseries Inverted (black background)\\
  T1 & 0.04 (0.01, 0.07) & 0.28 (0.21, 0.34)  & 0.53 (0.46, 0.6)\\
  T2 & 0.28 (0.21, 0.36) & 0.06 (0.03, 0.08)  & 0.12 (0.09, 0.15)\\
\end{tabular}
\end{table}

\begin{table}
\centering
\begin{tabular}{cccc}

    \multicolumn{4}{c}{\bfseries Primary Tumor (ccRCC Model)} \\
  \hline
  \bfseries Sequence & \bfseries Unprocessed  & \bfseries Inverted & \bfseries Inverted (black background)\\
  T1 & 0.04 (0.01, 0.07) & 0.14 (0.09, 0.20)  & 0.42 (0.35, 0.49)\\
  T2 & 0.12 (0.07, 0.16) & 0.01 (0.00, 0.02) & 0.02 (0.00, 0.05)\\
\end{tabular}
\end{table}

\fi

\begin{figure}
\centering
\includegraphics[width=1\linewidth]{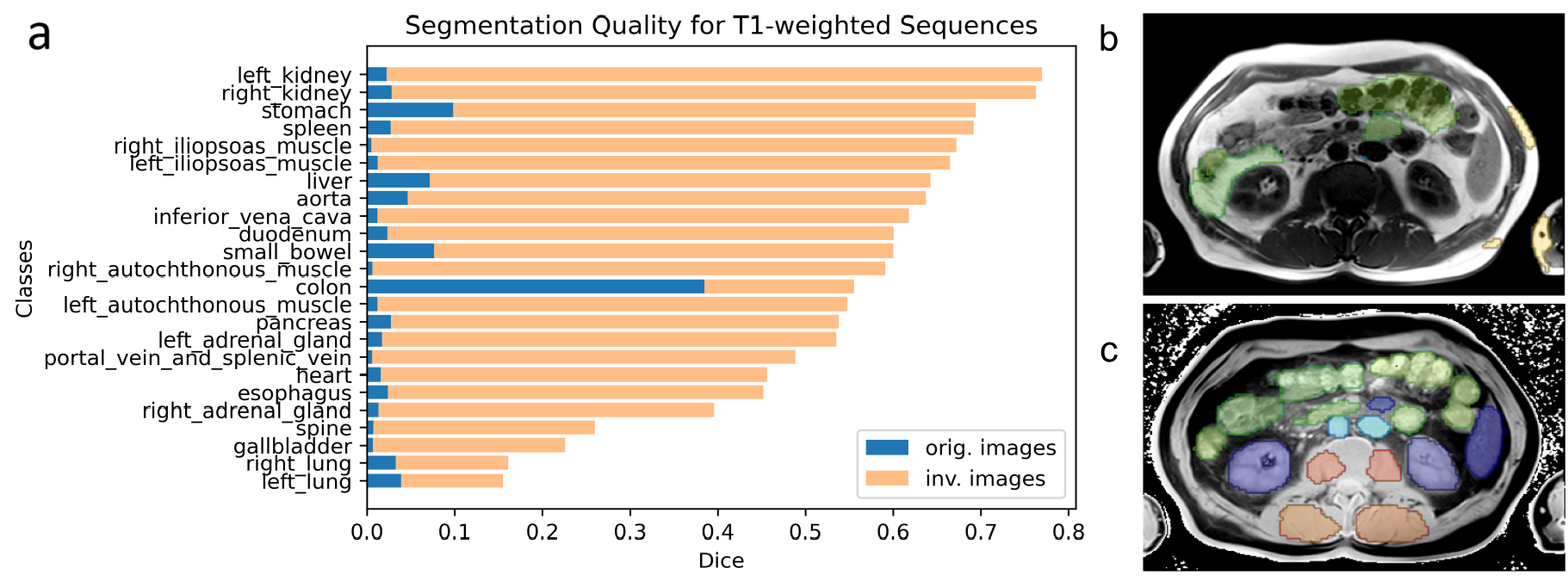}
\caption{(a) Mean DSC of segmentations by TotalSegmentator-fast for original and inverted T1-weighted MRI images. Segmentation before (b) and after (c) inversion.}
\label{fig:results}
\end{figure}

Without preprocessing, TotalSegmentator failed to detect any classes in the T1-weighted sequences, except for the colon (DSC=0.38). For T2-weighted sequences, TotalSegmentator could partially segment large organs but struggled to segment blood vessels and muscles. Inverting the sequences lead to improved segmentation quality for T1-weighted images but decreased the performance for T2-weighted images (Table \ref{tab:results}). Setting the background intensity to zero significantly enhanced the segmentation quality for T1-weighted sequences. We observed improvements across all classes, including small vessels, adrenal glands and the lungs (Figure \ref{fig:results}).

To assess the performance of the ccRCC model, we focused on three distinct areas: The tumor region, the non-tumor kidney area (referred to as $K_{tumor}$ ), and the tumor-free kidney (called $K_{clean}$).
On the CT validation fold the model achieved a DSC of 0.89 for $K_{tumor}$, 0.91 for $K_{clean}$ and 0.73 for the tumors. Contrary, for the unprocessed T1-weighted MRI no class could be correctly segmented (DSC <= 0.03). In the unprocessed T2-weighted images $K_{tumor}$ (DSC: 0.57) and $K_{clean}$ (DSC: 0.63) could be somewhat segmented but tumor segmentation remained unsuccessful (DSC: 0.11).

Adding the inversion step increased the segmentation accuracy for kidneys (DSC = 0.76 / 0.71) and tumors (DSC = 0.45) in the T1-weighted images but prevented segmentation of the T2-weighted images (DSC <= 0.10 for all classes).
For the specific case T1 inverted: tumors were correctly localized in 75 scans, incorrectly in 19 scans, and could not be detected in 6 scans. The tumors in these groups had a median volume of 35 cm$^3$, 23 cm$^3$ and 6 cm$^3$, respectively. Tumors below the median volume of 29 cm$^3$ were segmented with a DSC of 0.22 and tumors above the median volume with a DSC of 0.62. A paired two-sided t-test showed a significant correlation between tumor volume and DSC (p<0.001).

\section{Discussion}
Creating annotation from scratch is a work intensive and time-consuming process.
We showed that a simple image inversion technique can significantly improve the segmentation quality of CT segmentation models for T1-weighted MRI data. We demonstrated the feasibility for both a general multi-class and a specialized renal tumor model. Using this technique, we were able to localize and segment large ccRCC in T1-weighted MRI scans, using a model that was trained on CT data only. Our results underline that transferring available CT models, including pathology models, to the MRI domain can provide a viable efficiency boost in projects requiring large-scale pixel wise annotation.

The results emphasize the importance of color gradients for the tested model. In particular, the contrast between the participants and the background appears to be crucial. Consistently setting the background in an inverted image to black improved the segmentation of all classes, even for organs located in the center of the body.
Although the results were promising for T1-weighted images, we could not demonstrate improvements for T2-weighted images. This is likely due to the increased intensity of water in T2 images, which causes most organs to appear brighter than surrounding tissue. Additionally, the lungs were difficult to segment in all tested sequences, likely due to the air within the lobes.

For T1-weighted MRI, inversion improved segmentation of both tested models. Contrary to many abdominal organs with clear distinctive border, renal lesions can have similar attenuation to the kidneys \cite{de2017imaging}. Despite this, we were able to segment ccRCC in many instances. Segmentation quality greatly correlated with tumor volume, with larger tumors being localized and segmented much better. However, this could partially be a side-effect of using the DSC as a metric, which is more suitable for larger structures \cite{maier2024metrics}. 

\begin{figure}
\centering
\includegraphics[width=0.9\linewidth]{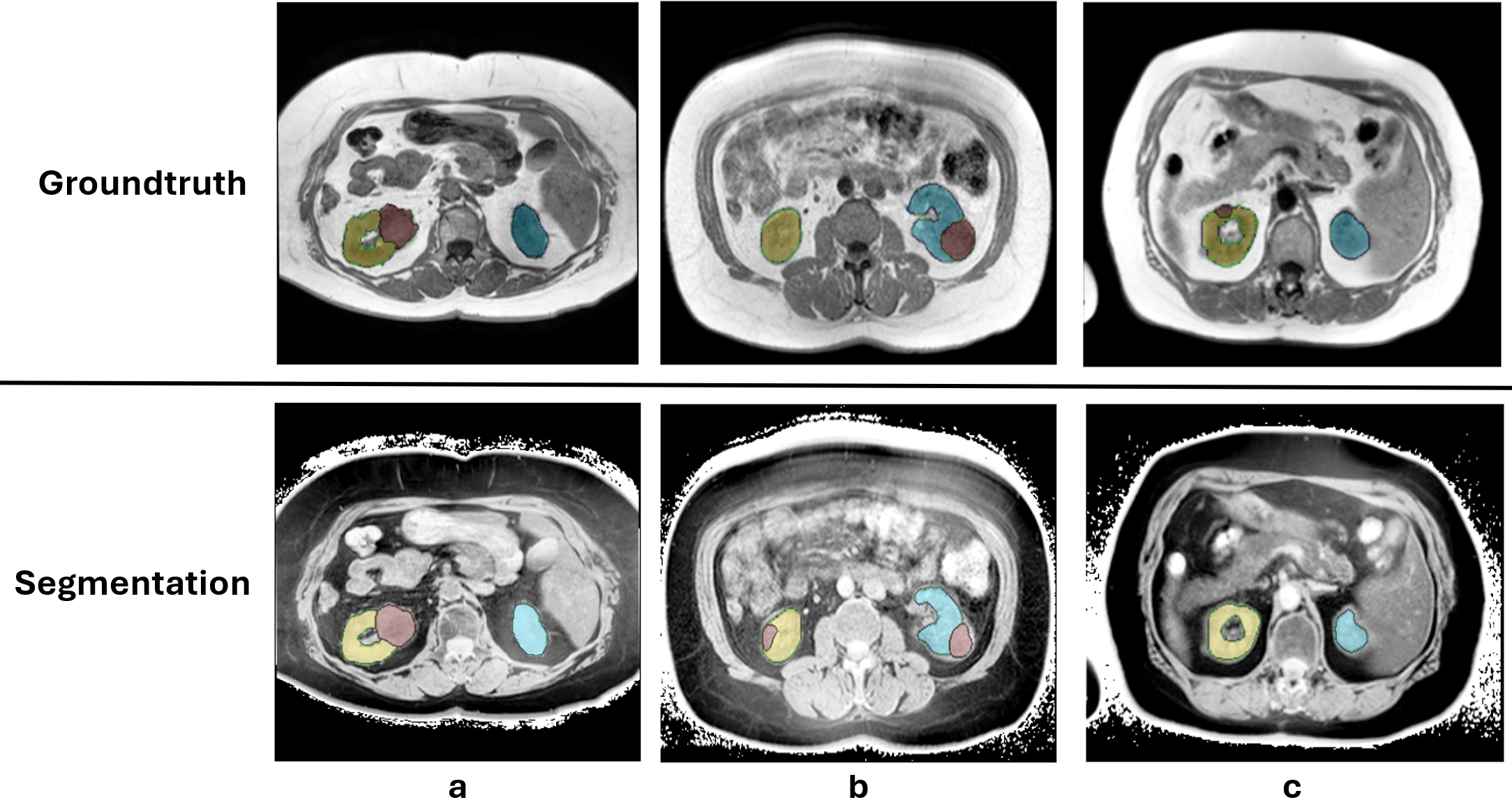}
\caption{Groundtruth and segmentations of clear cell kidney carcinoma in T1-weighted MRI. (a) Ideal case: Primary tumor and kidney are correctly separated (DSC: 0.94). (b) The model segments a possible secondary tumor that is not included in the groundtruth, resulting in a bad evaluation (DSC: 0.51). (c) Model fails to correctly delineate the tumor (DSC: 0.00). }
\label{fig:kidney_examples}
\end{figure}

Our study has acknowledged limitations. First, we created our groundtruth using both manual annotation and automatic segmentation. Doing this reduces the meaningfulness of the DSC, as we cannot assume the groundtruth to be completely correct. However, we consider it sufficient in the context of our study, as the goal is not 100\% accuracy, but rather creating pre-segmentations for a faster annotation process. 
Second, we focused on primary ccRCC and disregarded secondary tumors or metastasis. A segmentation of these would have been classified as incorrect by our evaluation pipeline, while a radiologist might have decided otherwise (Figure \ref{fig:kidney_examples}).

Ideally, our strategy should be used as the first step in an active learning framework, such as MONAI Label \cite{nath2020diminishing, DiazPinto2022monailabel}, for annotating a new structure. Inversion with subsequent segmentation can be quickly applied to a large dataset. Annotators can then assess the quality at a glance and focus on the best segmentations, while saving worse results for later. Given sufficient MRI sequences it may even be enough to train an nnU-Net without any new or little annotation effort \cite{philipp2024annotationefficient}. 

More sophisticated modality-transfer methods could potentially increase generalizability, however, implementing these complex models can be challenging and may be disproportionate for small-scale projects. For certain sequences, image inversion alone can be sufficient to achieve satisfactory results.

\section*{Acknowledgements}
The authors acknowledge the Scientific Computing of the IT Division at the Charité - Universitätsmedizin Berlin for providing computational resources that have contributed to the research results reported in this paper (\url{https://www.charite.de/en/research/research_support_services/research_infrastructure/science_it/#c30646061}).
This work was in large parts funded by the Wilhelm Sander Foundation.
Funded by the European Union. Views and opinions expressed are however those of the author(s) only and do not necessarily reflect those of the European Union or European Health and Digital Executive Agency (HADEA). Neither the European Union nor the granting authority can be held responsible for them.

\begin{figure}[H]
\includegraphics[width=0.3\linewidth, right]{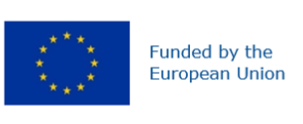}
\label{fig:eu_funding}
\end{figure}


\bibliographystyle{unsrtnat}
\bibliography{main.bib}  

\end{document}